\documentclass[runningheads]{llncs}
\usepackage[utf8]{inputenc}
\usepackage{graphicx}
\usepackage{tabularx}
\usepackage{comment}
\usepackage{booktabs}
\usepackage{caption}
\usepackage{enumerate}
\usepackage{url}
\usepackage[colorinlistoftodos]{todonotes}
\usepackage{paralist}
\usepackage{pgfplots}
\usepackage{pgf-pie}
\usepackage{cite}
\pgfplotsset{compat=1.18}

\newcommand\nk[1]{\textcolor{black}{#1}}

\makeatletter
\renewenvironment{description}%
               {\list{}{\leftmargin=15pt % <------- Adjust this length
                        \labelwidth\z@ \itemindent-\leftmargin
                        }}%
               {\endlist}
\makeatother
\begin{document}
%
%\title{Agentic Business Process Management: Practitioner's Perspectives on Agent Governance in Business Processes} 
\title{Agentic Business Process Management: Practitioner Perspectives on Agent Governance in Business Processes} 
%
%\titlerunning{Abbreviated paper title}
% If the paper title is too long for the running head, you can set
% an abbreviated paper title here
%
\author{Hoang Vu\inst{1}\orcidID{0009-0002-1656-7207}
\and Nataliia Klievtsova\inst{2}\orcidID{0009-0009-9010-2855}
\and Henrik Leopold\inst{3}\orcidID{0000-0003-4862-1829}
\and Stefanie Rinderle-Ma\inst{2}\orcidID{0000-0001-5656-6108}
\and Timotheus Kampik\inst{1,4}\orcidID{0000-0002-6458-2252}
}
\authorrunning{Vu et al.}
\titlerunning{Agentic Business Process Management}
% First names are abbreviated in the running head.
% If there are more than two authors, 'et al.' is used.
%
\institute{SAP, Berlin, Germany \\
\email{\{h.vu, timotheus.kampik\}@sap.com} \and Technical University of Munich, \\TUM School of Computation, Information and Technology, Garching, Germany \\
\email{\{nataliia.klievtsova,stefanie.rinderle-ma\}@tum.de}
\and Kühne Logistics University, Hamburg, Germany \\
\email{henrik.leopold@klu.org}
\and Umeå University, Umeå, Sweden}
\maketitle              % typeset the header of the contribution
\begin{abstract}
With the rise of generative AI, industry interest in software \emph{agents} is growing. Given the stochastic nature of generative AI-based agents, their effective and safe deployment in organizations requires robust governance, which can be facilitated by \emph{agentic} business process management.
However, given the nascence of this new-generation \emph{agent} notion, it is not clear what BPM practitioners consider to be an agent, and what benefits, risks and governance challenges they associate with agent deployments.
To investigate how organizations can effectively govern AI agents, we conducted a qualitative study involving semi-structured interviews with 22 BPM practitioners from diverse industries.
They anticipate that agents will enhance efficiency, improve data quality, ensure better compliance, and boost scalability through automation, while also cautioning against risks such as bias, over-reliance, cybersecurity threats, job displacement, and ambiguous decision-making. To address these challenges, the study presents six key recommendations for the responsible adoption of AI agents: define clear business goals, set legal and ethical guardrails, establish human-agent collaboration, customize agent behavior, manage risks, and ensure safe integration with fallback options. Additionally, the paper outlines actions to align traditional BPM with agentic AI, including balancing human and agent roles, redefining human involvement, adapting process structures, and introducing performance metrics. These insights provide a practical foundation for integrating AI agents into business processes while preserving oversight, flexibility, and trust.

\end{abstract}

\section{Introduction}
\label{sec:intro}

For over three decades, \textit{agents} have periodically surged in Business Process Management (BPM). The 1990s saw early excitement around goal-oriented software agents~\cite{DBLP:journals/ijcis/JenningsFJNOW96,DBLP:journals/sigmod/JenningsNF98}, followed by the rise of Robotic Process Automation (RPA) in the late 2010s~\cite{Flechsig2019RealizingTF,darkside,rpabpm}, promising efficiency gains for knowledge work. However, high initial expectations and maintenance costs led to mixed experiences in RPA adoption~\cite{hindle2017robotic}. 

Large Language Model (LLM)-based agents fuel another wave of optimism~\cite{DBLP:journals/tmis/DumasFLMMRACGFGRVW23,berti2024rethinkingprocessminingaibased}.
However, the stochastic nature of LLMs and related generative AI (genAI) technologies raises concerns, particularly regarding the need to govern the artifacts, decisions, and behaviors they produce. This prompts a critical question: how can BPM support the responsible integration of AI agents into organizations, ensuring both effective business outcomes and social accountability?

To explore this challenge from a practical perspective, we conducted interviews with BPM practitioners, capturing their insights on the potential implications of agents on organizations. The results reveal a dual landscape of opportunities and risks. On the one hand, practitioners anticipate benefits such as improved efficiency, predictive insights, and proactive decision-making. On the other hand, they express concerns about bias, over-reliance on automation, lack of transparency, and potential job displacement. This paper aims to address these concerns by providing recommendations practitioners should consider when introducing autonomous agents in business processes.

The rest of the paper is organized as follows. Section~\ref{sec:background} reviews the history and evolution of agents in BPM, RPA, generative AI in BPM and defines the concept of \emph{agentic BPM}. Section~\ref{sec:methodology} describes the qualitative research design. Section~\ref{sec:results} presents findings from the study with BPM practitioners on agents, while section~\ref{sec:recommendation} highlights the key recommendations proposed for AI agent adoption. Section~\ref{sec:limitation} and~\ref{sec:conclusion} discusses the limitation of the study and concludes by highlighting the need for further research into governance frameworks for agents within business processes.

\section{Background}
\label{sec:background}

In this section, we provide a brief overview of the history and evolution of agents in BPM. We highlight both the continuous research progress in areas applying agent technologies in business processes (Sect.~\ref{sec:agents-no-gen-ai}), as well as RPA (Sect.~\ref{sec:rpa}). Finally, we discuss genAI in BPM (Sect.~\ref{sec:agents-gen-ai}) and define agentic BPM as the governed introduction of AI agents into business processes (Sect.~\ref{sec:agentic-bpm}).

\subsection{Agents in Business Process Management}
\label{sec:agents-no-gen-ai}

The integration of agent technologies into business process execution and coordination began in the early 1990s. An agent is typically defined as a computer program that ``\textit{operates autonomously, perceives its environment, endures over time, adapts to change, and pursues goals}''~\cite[pp.21--22]{RN2020}\footnote{Definition of \emph{agents} vary; discussing these definitions is beyond the scope of this paper.}. In the context of Multi-Agent Systems (MAS)\footnote{We use the terms \emph{agents} and \emph{MAS} interchangeably, considering a collection of agents as a MAS.}, a business process can be viewed as a set of interacting agents coordinated by common organizational goals~\cite{DBLP:journals/aai/JenningsNOO00,DBLP:journals/sigmod/JenningsNF98}; accoridngly Agent-Oriented Programming (AOP) focuses, in its application proposals to BPM, primarily on business process execution~\cite{DBLP:journals/ijcis/JenningsFJNOW96}). Over the decades, researchers have applied MAS to BPM to create more dynamic, cross-organizational processes while maintaining overall control. However, early implementations often faced challenges, including weak system control and issues with trust and delegation~\cite{agentdis}. Nonetheless, such processes, as well as the MAS executing them, need to be managed to align with organizational and broader societal requirements.
%From a MAS perspective, the long-running line of research on \emph{normative MAS} studies this problem~\cite{chopra2018handbook,10.1145/1082473.1082575}.
%It has some overlaps with BPM, e.g., in works that propose the application of deontic logic (formalizing notions such as obligation, prohibition and permission) to business process compliance~\cite{10.1145/2514601.2514608}.
%Modern organizational processes are increasingly decentralized, dynamic, complex, and knowledge-intensive, operating as MAS where both humans and software systems act autonomously. Consequently, BPM frameworks must not only integrate new agent-based technologies but also evolve to govern the autonomy of these agents effectively. This challenge is underscored in both research on web-scale systems \cite{10.1145/3507910} and more focused studies addressing the management and governance of agent autonomy in business processes \cite{Marcio2004AddressingAA}.

In process modeling, \emph{process choreographies} describe how independent agents coordinate—following the BPMN standard~\cite{omg2011bpmn}—by exchanging messages in a defined sequence and under specific conditions to accomplish a shared goal~\cite{decker2009design}.
Similarly, subject-oriented BPM shifts the focus from traditional process flows to subjects and their interactions, where subjects (e.g., humans, software, or agents) autonomously manage their tasks and decisions~\cite{SOBPM}. With the rise of blockchain technologies, research further explored how networks of autonomous agents could underpin data sharing and decision logic through smart contracts~\cite{DBLP:journals/tmis/MendlingWABCDDC18,DBLP:conf/bpm/LadleifWW19}. Although blockchain-based BPM gained traction as a potential future framework, it did not move autonomous agents into mainstream BPM practice.
%More recently, the need for Agent-Based Modeling and Simulation (ABMS) in BPM has been explored with increased attention~\cite{agentsim}.

In process analysis, agent-based methods have been developed for tasks such as process mining and simulation~\cite{agentmining,sulis2022agent}, including data-driven simulations that utilize event logs~\cite{AgentSimulator}. \emph{Agent System Mining} (ASM) combines process mining with agent-based modeling to derive MAS models from real-world data~\cite{DBLP:conf/bpm/TourPKS23,DBLP:journals/access/TourPK21,DBLP:conf/er/ShenPLK24,DBLP:journals/access/TourPK21}.
\emph{Reinforcement learning} approaches have also been applied, enabling agents to test and refine different process variants through trial and error~\cite{aaronabtesting,DBLP:journals/is/SatyalWPCM19}.

\subsection{Robotic Process Automation}
\label{sec:rpa}

\nk{While agents and MAS address complex decision-making and coordination challenges, RPA aims to efficiently automate simple, rule-based tasks on a large scale. Such tasks represent a considerable amount of routine business activities~\cite{Flechsig2019RealizingTF}.}

\nk{Since 2015, research on RPA has highlighted numerous examples of how business process automation can significantly enhance performance~\cite{rpabpm}. When combined with BPM, RPA offers several advantages, including scalability, improved process accuracy, greater transparency and traceability, cost savings, and increased job satisfaction~\cite{Flechsig2019RealizingTF,darkside}. However, successfully applying RPA remains challenging, with 30-50\% of initial projects failing during implementation~\cite{hindle2017robotic}.}

%\nk{The primary causes of RPA failures result from technical limitations. These include scalability challenges, the absence of error recognition mechanisms, inflexibility in UI integration, and data security risks that endanger sensitive business processes, making automation more complex. Additionally, incorrect cost estimation, as well as issues related to maintenance, governance, and reliance on human expertise, contribute to project failures~\cite{rpachallenge,darkside,rpaproblem}.}

While the primary causes of RPA failures result from technical limitations, there is also the problem of resistance against automation. Automated decision-making must comply with regulations, as the absence of human oversight can lead to legal and ethical risks. Automation may also create a black box effect, making it difficult for employees to understand the logic behind automated decisions~\cite{rpabpm}. Therefore, it is crucial to consider the effects of human-automation interaction alongside the technical aspects of introducing new technology~\cite{vu2023towards}.

Even if successful, RPA is not suitable for full-scale business process automation as it focuses on automating repetitive tasks rather than orchestrating end-to-end processes~\cite{rpaproblem}. Since workflow automation and Integration platform as a service (iPaaS) are better suited for human-in-the-loop and API integration scenarios respectively, the automation use cases for RPA are limited~\cite{vu2023business}.

%\nk{These limitations, combined with the absence of dynamic decision-making capabilities that ensure flexibility and adaptability, emphasize the need for an innovative process automation approach involving human-like intelligence and genAI-based agents~\cite{Ye2023ProAgentFR}.}

\subsection{Generative AI in Business Process Management}
\label{sec:agents-gen-ai}

With the rise of LLMs, BPM researchers have begun exploring how these powerful tools can be integrated into business process management. Several studies~\cite{bpmllm1,bpmllm2,bpmllm3} have examined all stages of the BPM lifecycle to identify areas where LLMs can add value. However, much of the current literature focuses on using LLMs from a conversational perspective—for tasks such as process understanding, information retrieval, and decision support—primarily addressing specific phases like \emph{discovery}, \emph{analysis}, and \emph{monitoring}~\cite{Landscape}. While these approaches may treat LLMs as agents in a limited sense, they largely automate isolated tasks, leaving final decision-making and broader, goal-based autonomy to human users.

In contrast, long-term visions of generative AI consider AI-based systems not merely as tools for task automation but as intelligent systems capable of adaptability, self-improvement, and bounded autonomy. Recent works~\cite{DBLP:journals/tmis/DumasFLMMRACGFGRVW23,largeprocess,bpmai} present AI-Augmented BPM Systems (ABPMSs) that enhance process execution, analysis, and optimization within structured workflows by supporting human decision-making rather than replacing it. Notably, ABPMS are envisioned to support BPM holistically~\cite{DBLP:journals/tmis/DumasFLMMRACGFGRVW23}, while emerging research explores the application of LLM agents to process mining~\cite{berti2024rethinkingprocessminingaibased}. Similarly, Large Process Models (LPMs) adopt a neuro-symbolic approach by integrating fine-tuned LLMs with traditional symbolic systems for context-aware process design and analysis.

Despite these advances, a critical open question remains: How can BPM best support the management of increasingly autonomous software agents deployed in organizations, and what shifts in perspective are necessary to achieve this?

\subsection{Agentic Business Process Management}
\label{sec:agentic-bpm}

\begin{figure}[tb]
    \centering
    \includegraphics[width=0.95\textwidth]{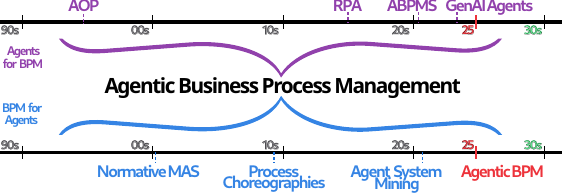}
    \caption{Emergence of Abstractions and Technologies for agentic Business Process Management over Time: \emph{Agents for BPM} versus \emph{BPM for Agents}.}
    \label{fig:abpm}
\end{figure}

Building on the research history of agents in BPM, RPA and generative AI in BPM (see Sect.~\ref{sec:background}),  we introduce the concept of \emph{agentic Business Process Management} (see Fig.~\ref{fig:abpm}) to address this key question.
Figure~\ref{fig:abpm} presents a historical overview that situates agentic BPM within the evolving relationship between agent systems and business process management. Rather than serving merely as a conceptual map, the timeline illustrates how the field has progressed—from leveraging agents to support BPM tasks (“Agents for BPM”) toward rethinking BPM as a means to govern and coordinate agent-based systems (“BPM for Agents”). This distinction is crucial: while the former reflects the application of agents in traditional BPM contexts, the latter positions BPM as a discipline that adapts and evolves to meet the challenges posed by autonomous, self-adaptive multi-agent systems (MAS). Highlighting this transition helps clarify why a dedicated framework such as ABPM is necessary—not just to harness the capabilities of AI agents but also to ensure their responsible and effective integration into organizations. Our definition of ABPM remains technology-agnostic, draws on established BPM and AI terminology, and is intended to contribute to the broader evolution of BPM as a discipline for agent governance.

\begin{definition}[Agentic Business Process Management]\label{concept1} \normalfont Agentic Business Process Management (ABPM) describes \emph{i)} the deployment and execution of autonomous software agents to achieve business process goals and \emph{ii)} the application of agent-based abstractions for the process-oriented design and analysis of autonomous software agents.
\end{definition}

Definition \ref{concept1} makes use of several key concepts, all of which are well-established in the BPM and AI literature.
\begin{itemize}
    \item An \emph{(autonomous) software agent} is a computer program that ``operates autonomously, perceives the environment, persists over a prolonged time period, adapts to change, and creates and pursues goals''~\cite[pp.21]{RN2020}.
    \item A \emph{process goal} operationalizes the objective that an organization strives to achieve with the corresponding business process, which---in turn--can be defined as ``a set of activities that are performed in coordination in an organizational and technical environment'', to ``jointly realize a business goal''~\cite{Weske2019}.
    \item An \emph{agent-based abstraction} is a conceptual model explicitly featuring notions of (software or human) agents.
    \item \emph{Process-oriented design and analysis} describes the modeling business processes, as well as the drawing of inferences from data generated during process execution. 
\end{itemize}

These clarifications build on textbook definitions of agents and business processes. We have deliberately omitted common and nuanced discussions on what constitutes an agent (cf. \cite[Chapter 1]{10.5555/305606}). We do not assume that agents are perfectly rational and expect them to always act in a way that achieves the best expected outcome \cite[pp.22]{RN2020}. Instead, our only technological commitment is that agents are software-based, supporting self-directed work that extends beyond human capabilities.

Our definition of agentic Business Process Management (ABPM) spans the key phases of the BPM lifecycle by addressing both deployment/execution and design/analysis. Although the discovery phase often overlaps with analysis in practice, our conceptual and technology-neutral definition applies BPM principles to multi-agent systems (MAS).

\section{Methodology}
\label{sec:methodology}

To explore how AI agents can be safely adopted in organizations and what the implications are for BPM, we utilized a qualitative research design and conducted in-depth interviews with experienced BPM practitioners. The insights gained reveal both the advantages and challenges associated with deploying AI agents, enabling us to identify key implications and formulate targeted actions to address critical governance and management concerns. \\

\noindent \textbf{Terminology.} Several process management software vendors, including UiPath, Salesforce, IBM, and Workday, use the term \emph{agentic AI} to describe the interplay between genAI and agents.
Thus, we leveraged the term during the interviews as a starting point to discuss the impact of agents on process management, expecting it to resonate effectively with the interviewees. \\

\noindent \textbf{Research design.} Our study leverages a qualitative research design to examine the perspectives of professionals in business process management and automation, focusing on the integration and impact of agentic AI. 
It investigates the participants' understanding, expectations, and concerns related to agentic AI's autonomy, adaptability, human collaboration, and governance. Alongside discussing the characteristics of agentic AI, the study places greater emphasis on its impact on practitioners' organizations. Therefore, the study also explores to what extent agent governance is required to address management needs that arise from greater software autonomy.

To systematically examine the perceptions and attitudes towards agentic AI, a qualitative content analysis approach \cite{Mayring2019} was followed. Semi-structured interviews were conducted with professionals across various industries, focusing on their understanding, expectations, and concerns regarding autonomy, adaptability, human collaboration, and governance of agentic AI. The process began with transcription, familiarization with the interview data and analysis using a deductive-inductive coding approach. Initial deductive categories were derived from the research questions, while additional inductive subcategories emerged from the data, allowing themes to derive directly from the participants' responses. 

\begin{figure}[bth]
    \centering
    \includegraphics[width=\textwidth]{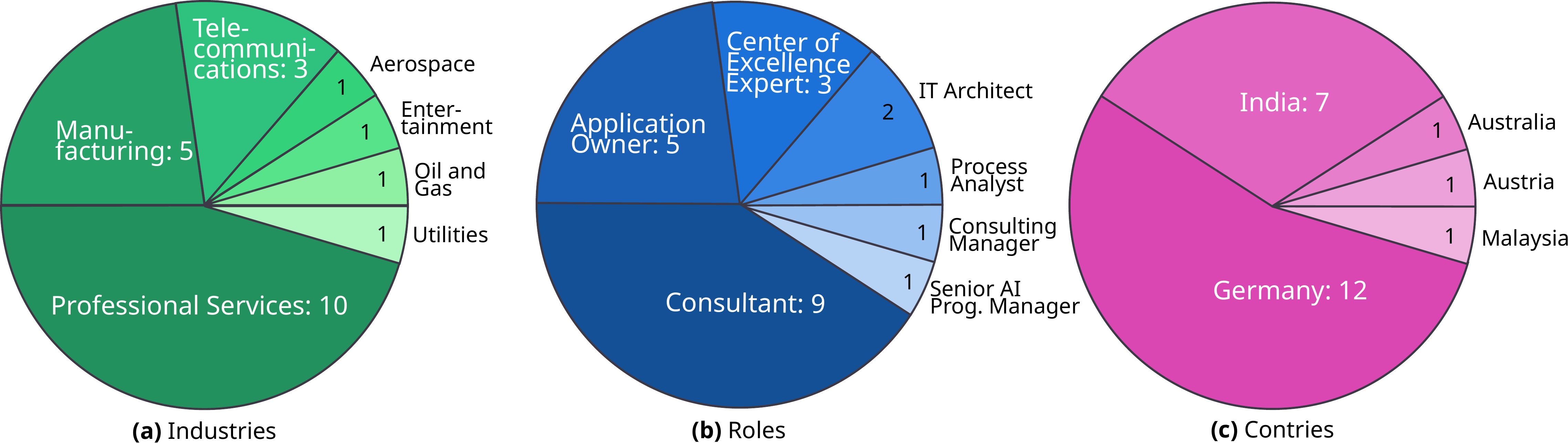}
    \caption{Demographic Distribution across Participants}
    \label{fig:demographic}
\end{figure}

We began by reviewing all the interview data to understand the context and identify recurring themes. Next, we applied open coding by labeling segments of text with basic descriptive codes (e.g., benefits, risks, and governance challenges related to agentic AI). We then grouped similar codes into broader categories during axial coding (e.g. "automation potential" and "workflow optimization" to "benefits"), revealing patterns and common themes across the interviews. Finally, we refined and integrated these themes into a coherent framework during selective coding, ensuring that key insights were clearly defined.
A total of 22 participants with experience with automation and AI technologies from various industries and roles were interviewed by the first author. 
In Figure~\ref{fig:demographic}, an overview of the participants demographic characteristics is provided, including their professional roles and the industries and countries in which they are employed.
Sessions lasted 60 to 90 minutes, with some conducted in pairs or groups due to accessibility constraints.

\section{Results}
\label{sec:results}
%%%%%%%%%%%%%%%%%%%%%%%%%%%%%%%%%%%%%%
The interview results highlight various themes mentioned by the interviewees, which are discussed below. \\

\noindent \textbf{Understanding.} Among the 22 participants, 10 were familiar with the term agentic AI, describing it as a self-learning technology that operates autonomously and adapts to its environment. Some viewed it as an evolution of RPA, overcoming technical limitations with AI. Others associated it with a digital assistant or an orchestration layer that coordinates tasks across specialized agents. In this context, a participant highlights:
\begin{quote}
"[Agentic AI is an agent] that can do significantly more than dumb botting, where I always have to tell it everything exactly and have precisely structured it, and it really only helps where I have very clearly structured, stable processes."
\end{quote}

\noindent Importantly, none of the participants reported to have actual practical experience with agentic AI, i.e., the assessments participants provide is based on expectations, often against the backdrop of existing agent-like automation technologies such as RPA. \\ 

\begin{figure}[tbh]
    \centering
    \includegraphics[width=\textwidth]{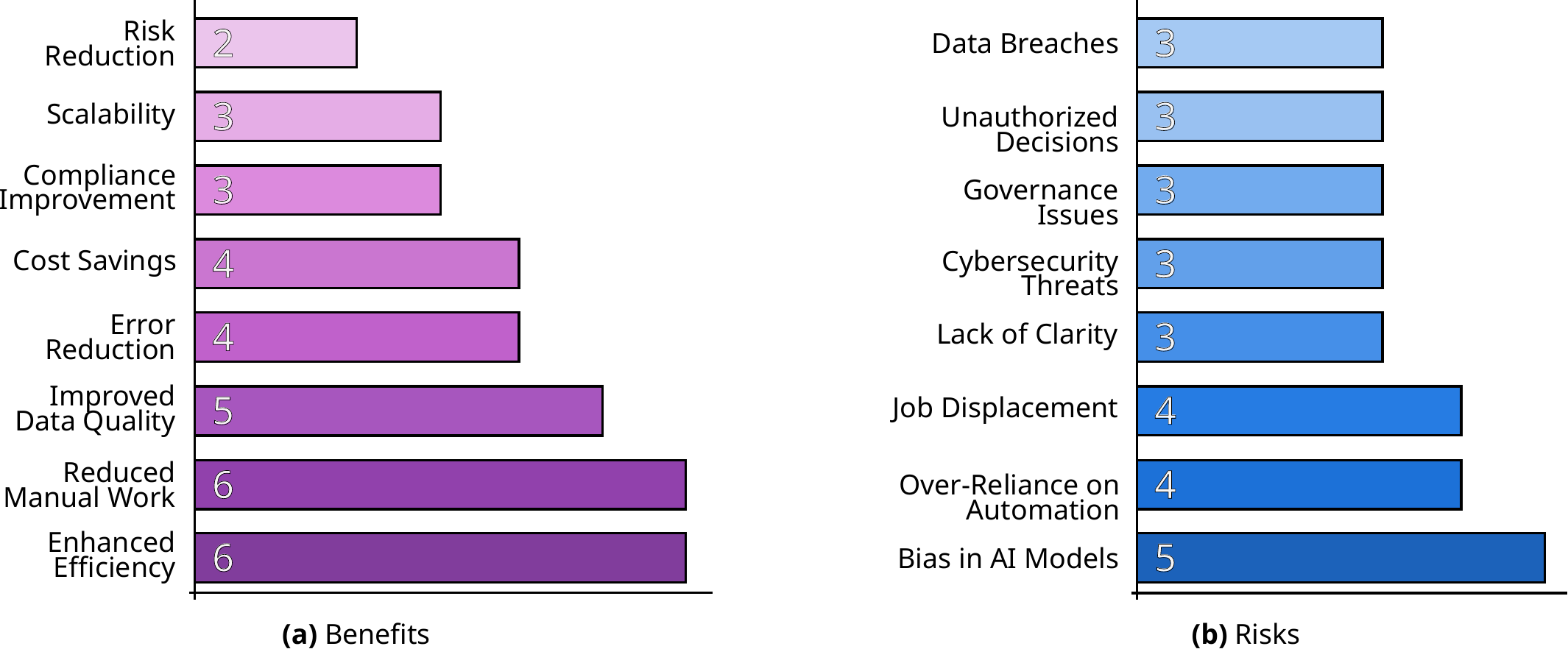}
    \caption{Agentic AI among Participants: Expected Benefits and Risks.}
    \label{fig:benefits-risks}
\end{figure}

\noindent \textbf{Benefits.} 
Participants highlighted several potential benefits of integrating agentic AI into organizational processes (see Figure~\ref{fig:benefits-risks} (a)). The technology is expected to enhance efficiency by automating routine tasks, streamlining operations, and reducing errors, allowing employees to focus on higher-value work and achieving time and cost savings. It is also anticipated to improve data quality, ensuring consistent and accurate handling of information. Better compliance was another potential benefit, as agentic AI could monitor regulations and enforce standards automatically. Scalability was frequently noted, enabling businesses to handle larger workloads without proportional increases in staffing. Additionally, it is believed to democratize process data, making it more accessible and usable. In this context, a participant highlights: 
\begin{quote}
"More employees could initiate changes or optimizations if the software supports them directly, reducing dependency on specialized roles."
\end{quote}

\noindent \textbf{Risks.} 
Despite its potential, implementing agentic AI carries risks (see Figure~\ref{fig:benefits-risks} (b)), such as bias from flawed training data, over-reliance leading to diminished human judgment, and lack of transparency in AI-driven decisions. Participants also raised concerns about cybersecurity threats, job displacement, and unauthorized decision-making, stressing the need for adequate human oversight to prevent unintended consequences. In this context, a participant highlights the cultural aspect:
\begin{quote}
"It's a cultural thing to be able to accept autonomy and decision making being taken away from a human."
\end{quote}

\begin{figure}
    \centering
    \includegraphics[width=\textwidth]{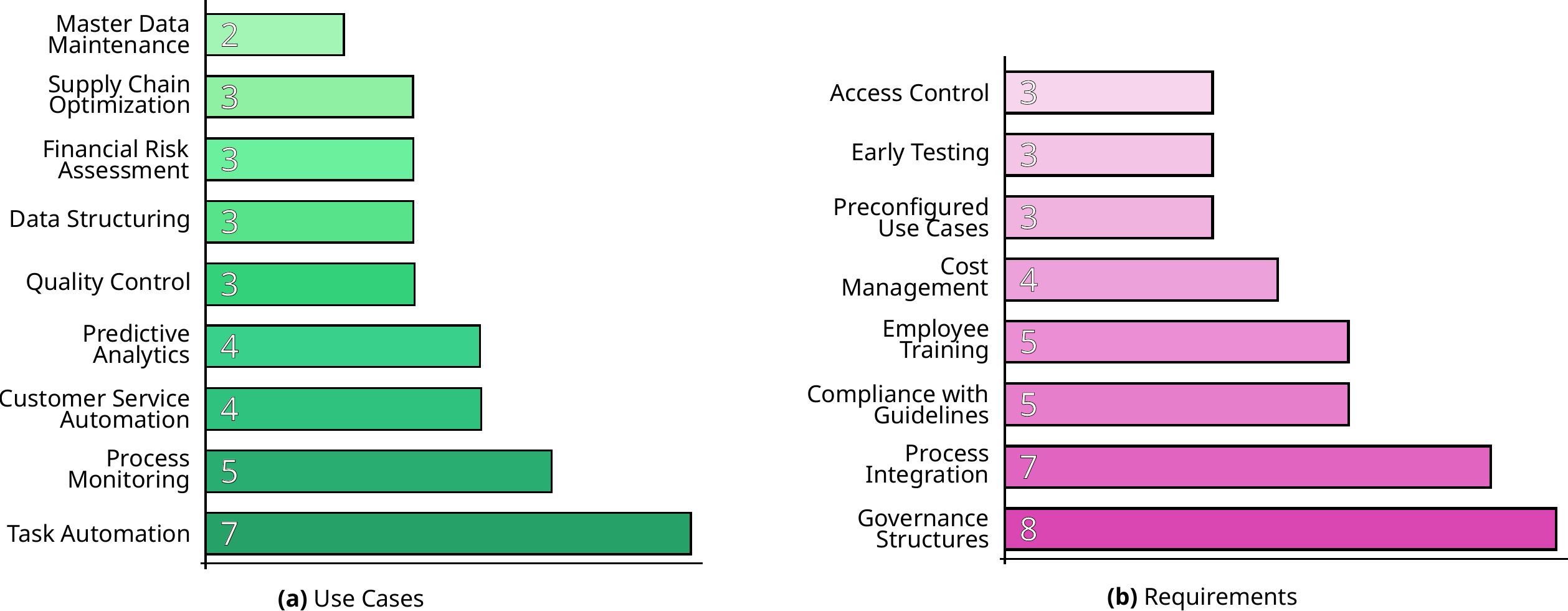}
    \caption{Agentic AI among Participants: Use Cases and Roll-out Requirements.}
    \label{fig:cases-req}
\end{figure}

\noindent \textbf{Use cases.} Participants identified several potential use cases for agentic AI (see Fig.~\ref{fig:cases-req}  (a)). Key applications include process monitoring to detect inefficiencies and suggest improvements, and predictive analytics to forecast trends and provide actionable insights. Task automation was another major focus, with agentic AI handling routine activities like data entry and document processing. Specific tasks include master data maintenance, user administration, root cause analysis, and decision support through dashboards. A participant highlights:
\begin{quote}
"Processes often get stuck due to errors in master data, such as mismatched product codes or pricing issues. [Agentic] AI could analyze and fix these autonomously."
\end{quote} 
Agentic AI could enhance customer service, while supply chain optimization benefits from agentic AI managing inventory and predicting demand. In finance, agentic AI could aid in fraud detection and transaction monitoring. Additionally, agentic AI could structure unstructured datasets, improving accessibility and analysis. \\

\noindent  \textbf{Requirements.}
Participants stressed the need for clear rules and guidelines to ensure agentic AI operates ethically and transparently. They highlighted the importance of audit logs, data retention, transparency, robust security, and adherence to corporate policies and regulations. In this context, a participant highlights:
\begin{quote}
"[The agent] would basically replace an FTE\footnote{\emph{Full-Time Equivalent}, i.e., a common unit to measure work in terms of person efforts.}, let's just put it that way; you also have to provide it with the same framework that the employee would be confronted with because what would the employee do if they encounter difficulties?"
\end{quote}
Defining roles, responsibilities, and limitations for agentic AI is crucial, along with compliance with data protection laws and risk management frameworks. Seamless integration with existing processes prevents disruptions, while comprehensive employee training ensures effective use of agentic AI. Managing costs, including setup, maintenance, and upgrades, was also highlighted as essential. Preconfigured use cases and applications were recommended to build trust by demonstrating agentic AI's capabilities and delivering clear benefits. Figure~\ref{fig:cases-req} (b) gives an overview of requirements that practitioners highlighted.

\noindent \textbf{Autonomy and adaptability.}
Participants stressed the importance of managing agentic AI’s autonomy to maintain accountability, reliability, and safety. While agentic AI can independently handle low-risk tasks like data preparation, human oversight is essential for critical decisions or significant changes. Configurable autonomy was recommended, granting agentic AI flexibility in low-risk areas while restricting it in high-risk scenarios, such as system changes or external communications. A gradual approach was also suggested, starting with simple tasks and increasing autonomy as trust in the system grows. In this regard, a participant highlights:
\begin{quote}
"It shouldn’t make changes in source systems or install new apps autonomously. That crosses the line because those areas are managed by different teams and require coordination."
\end{quote}
Adaptability is a key strength of agentic AI when operating within clear boundaries, but concerns about trust and consistency highlight the need for human oversight to monitor and guide its adaptations. Transparency was highlighted as essential, with participants calling for clear documentation of all AI decisions and actions. In this context, a participant mentions:
\begin{quote}
"If it’s routine, the decision itself should be documented. The more complex the task, the more I want to see how the process was developed."
\end{quote}
This would include high-level summaries for management and detailed logs for technical teams to ensure traceability and allow for changes to be reverted if necessary. Participants acknowledged agentic AI's potential to enhance workflows and efficiency but emphasized the need to operate within predefined rules to prevent unintended outcomes. Agents should support, not replace, humans in tasks requiring complex judgment. Human validation and oversight remain crucial for critical scenarios. \\

\begin{figure}
    \centering
    \includegraphics[width=0.7\textwidth]{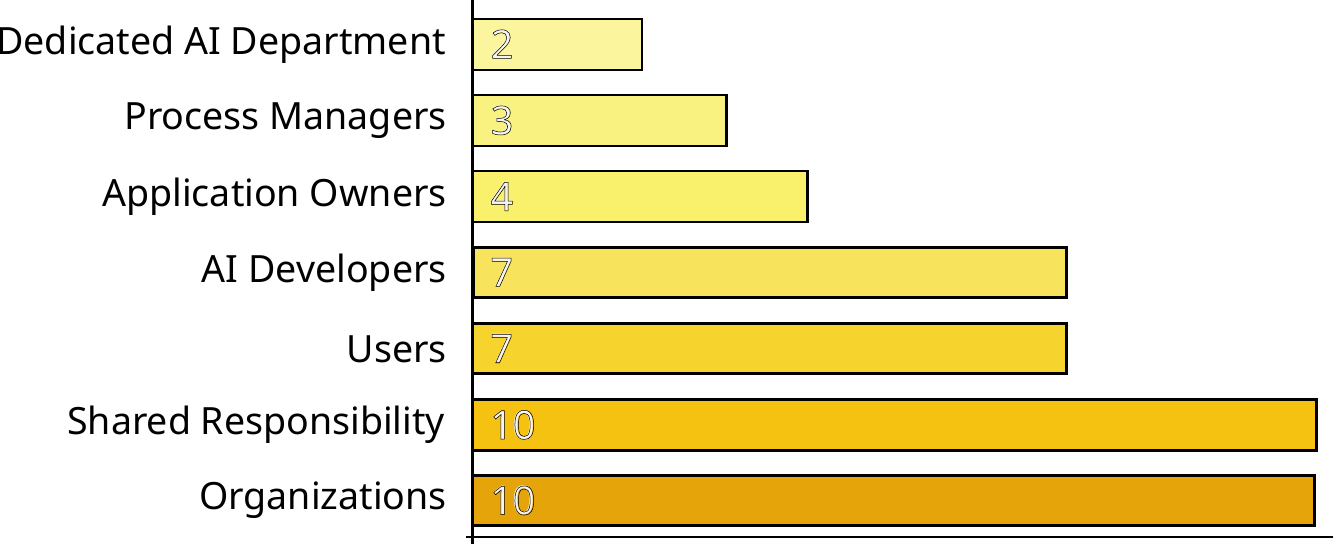}
    \caption{Agentic AI among Participants: Responsibility for Failures.}
    \label{fig:responsibility}
\end{figure}

\noindent  \textbf{Human involvement and responsibility.}
Participants agreed that although agentic AI can autonomously handle low-stakes decisions, it should serve as a decision-support tool for complex process scenarios, ensuring human involvement to maintain accuracy and accountability. Its role is to provide clear analyses, suggest actions, and outline impacts, enabling well-informed human decisions. Some proposed methods like briefings or notifications to engage humans in urgent situations, viewing agentic AI as a collaborative assistant offering insights and recommendations. In this regard, a participant underlines:
\begin{quote}
"For complex decisions, the [agentic] AI should provide full context; how it arrived at the decision, what data it used, and the potential consequences; just like how humans consult colleagues for advice."
\end{quote}
Responsibility for agentic AI failures was seen as a shared effort among organizations, developers, and business leaders. Organizations deploying agentic AI bear ultimate responsibility, particularly by providing oversight in critical tasks to validate agentic AI decisions. Developers play a key role in building reliable systems, while organizations establish robust oversight mechanisms. Key users and application owners were identified as primary contacts for resolving issues, with some suggesting dedicated teams to manage and ensure accountability for agentic AI. Figure~\ref{fig:responsibility} summarizes responsibilities of agentic AI failure that practitioners highlighted.

\vspace{-20pt}
%%%%%%%%%%%%%%%%%%%%%%%%%%%%%%%%%%%%
\section{Recommendations for ABPM Adoption}
\label{sec:recommendation}
The findings reveal that organizations require a clear and structured management framework to successfully integrate AI agents into their business processes while ensuring that human actors remain informed, involved, and empowered. Insights from the expert interviews led to a set of practical recommendations and actions that organizations should consider during implementation. These span six critical focus areas: business context, operational guardrails, human-agent collaboration, customization, risk management, and adoption.

\begin{description}
    \item[Business context.] Organizations should define specific goals for the use of AI agents, clarify the expected benefits, and assess the short- and long-term costs associated with their deployment and maintenance.
    \item[Guardrails.] Organizations must ensure that agents operate within the boundaries of legal, ethical, and organizational rules, while considering the context and environment in which the agents function.
    \item[Human-agent collaboration.] Clear roles and responsibilities should be established to enable effective collaboration between humans and software agents, including identifying situations where human intervention is necessary in case of agent failure.
    \item[Customization.] Organizations should tailor agents to meet their specific needs, determining factors such as the level of autonomy granted to agents, how human oversight and review are conducted, the extent of documentation required, and the desired level of reasoning and transparency in agent operations.
    \item[Risk management.] Organizations must implement safety measures to monitor performance and address potential risks, including preventing undesired adaptability or evolution of the agents.
    \item[Adoption.] Organizations must ensure the seamless and responsible integration of agents into existing business processes and systems, while implementing fallback mechanisms to enable a safe reversion if necessary.
\end{description}

In addition, the implications for BPM need to be considered. To bridge traditional BPM (focused on structure, control, and standardization) with agentic AI (emphasizing autonomy, adaptability, and self-learning), following actions are proposed to ensure their effective coexistence and alignment.
\begin{description}
    \item[Human-agent balance.] Organizations should strive to design workflows that strategically leverage both AI capabilities and human expertise. This balance is essential for optimizing efficiency while preserving human judgment and creativity.
    \item[Human involvement.] As intelligent agents become integral to business processes, organizations must clarify whether human actors will collaborate with agents as peers or take on supervisory roles with selective intervention.
    \item[Process structures.] To fully realize the potential of self-adaptive agents, organizations should consider evolving beyond static process models and enabling more flexible, dynamic structures that can adjust in real time based on agent-driven insights.
    \item[Performance metrics.] Effective integration of agentic AI requires clear metrics and performance indicators. Organizations should develop governance frameworks that track the impact of agents on process efficiency, decision quality, and broader organizational outcomes.
\end{description}

\section{Limitations and Future Work} 
\label{sec:limitation}
Several limitations encountered during this study must be acknowledged. First, the sample was relatively small and predominantly drawn from experts at selected multinational organizations, which may not fully represent the diversity of experiences across various industries or among smaller enterprises. Second, the concept of agentic AI remains in its formative stage, as participants employed varied terminology and frameworks, making it difficult to develop a standardized definition. Consequently, the limited adoption and practical experience with this technology constrain the interpretation of the interview results. %Moreover, although the interviews provided rich qualitative detail, their design limits the generalization of the findings.

Against this background, future research should focus on refining and empirically validating the conceptual frameworks for agentic AI, with particular emphasis on its ethical, legal, and technical implications. Additionally, quantitative studies and broader analyses across different industries are needed to assess the wider impact and evolution of agentic AI within various organizational contexts, and to develop robust governance structures that ensure transparent and accountable decision-making. Finally, future research should also examine how the governance of agentic AI with its core characteristics of autonomy and adaptability should differ from the current frameworks used for established technologies such as Robotic Process Automation.

\section{Conclusion}
\label{sec:conclusion}
The rise of generative AI has sparked growing interest in its integration with software agents, giving rise to the concept of \emph{agentic AI}. While this emerging technology holds promise, its impact on business processes and the requirements for its effective management remain insufficiently understood. This paper addresses this gap by identifying key recommendations for organizations seeking to implement AI agents responsibly.

Interviews with BPM practitioners reveal that next-generation software agents can enhance efficiency by automating routine tasks and allowing humans to focus on higher-value activities. At the same time, practitioners express concerns about risks such as bias, over-reliance, job displacement, and a lack of transparency. These concerns underscore the need for a clear governance framework to guide agent deployment.

To support informed, involved, and empowered human participation, the study outlines six critical areas for organizational focus: clarifying business objectives, setting legal and ethical guardrails, enabling effective human-agent collaboration, tailoring agent behavior, managing operational risks, and ensuring safe, seamless adoption.

Beyond implementation, the findings have broader implications for Business Process Management. Agentic AI introduces new dynamics—autonomy, adaptability, and self-learning—that challenge traditional BPM principles centered on structure, control, and standardization. To align these approaches, organizations should take strategic steps: design workflows that balance human and agent input, redefine human roles as collaborators or supervisors, evolve static process models to support adaptive agent behavior, and implement robust metrics and governance frameworks to monitor performance and outcomes.

Together, these insights offer a practical roadmap for the responsible integration of AI agents. They help organizations harness the potential of agentic AI while safeguarding process integrity, organizational control, and stakeholder trust.

%This emphasizes the importance of developing a methodology that ensures ethical use of agents through strong governance and active human oversight. The key requirements for such a framework can be broadly categorized into two foundational pillars. First, it aligns the business context, guardrails, and human-agent collaboration by setting clear goals, establishing strict operational boundaries, and defining roles for effective joint functioning. Second, it emphasizes customization, risk management, and adoption by tailoring agent functionalities to specific needs, implementing safety measures, and ensuring seamless integration with fallback mechanisms.

%The initial qualitative study presented in this paper highlights the need for broader research. Future studies should include participants with varied backgrounds, monitor the actual adoption of new generations of software agents (such as LLM-based agents), and assess the value of a defined agentic BPM framework. It should also investigate the application of agents in specific industries and develop strategies for its responsible adoption, balancing expected performance benefits with requirements regarding accountability and long-term system maintainability.

%%%%%%%%%%%%%%%%%%%%%%%%%%%%%%%%%%%%

%%%%%%%%%%%%%%%%%%%%%%%
%\section{Conclusions}
%\label{sec:conclusions}
%%%%%%%%%%%%%%%%%%%%%%%

\end{document}